\documentclass[%
preprint,
showpacs,preprintnumbers,
showkeys,
 amsmath,amssymb,
prd
]{revtex4-1}

\usepackage[T1]{fontenc} 
\usepackage{subcaption}
\usepackage{amssymb}
\usepackage{hyperref}
\usepackage{bm}
\usepackage{verbatim}
\usepackage{setspace}
\usepackage{float}
\usepackage[usenames,dvipsnames]{xcolor}
\usepackage[colorinlistoftodos]{todonotes}
\usepackage[normalem]{ulem}

\usepackage[compat=1.1.0]{tikz-feynhand}
\usepackage{tikz-feynman}
\tikzfeynmanset{compat=1.1.0}
\usepackage{feynmp}
\usepackage{tikzsymbols}
\usepackage{array}
\usepackage{pifont} 
\usepackage{bbold}
\usepackage{pifont}
\usepackage{array}
\newcolumntype{L}[1]{>{\raggedright\let\newline\\\arraybackslash\hspace{0pt}}m{#1}}
\newcolumntype{C}[1]{>{\centering\let\newline\\\arraybackslash\hspace{0pt}}m{#1}}
\newcolumntype{R}[1]{>{\raggedleft\let\newline\\\arraybackslash\hspace{0pt}}m{#1}}

\definecolor{titlecolor}{rgb}{0.6,0,0}
\definecolor{mightnightblue}{RGB}{25,25,112}
\definecolor{brown}{rgb}{0.59, 0.29, 0.0}
\definecolor{darkred}{rgb}{0.6,0,0}
\definecolor{linkcolor}{rgb}{0,0,0.5}

\usepackage{natbib}

\begin{document} 
\preprint{APCTP Pre2021 - 029, CTP-SCU/2021033}
\preprint{Prepared for submission to Phys.~Rev.~D}
\title{\boldmath \color{BrickRed}Dirac Radiative Neutrino Mass with Modular Symmetry and Leptogenesis}
\author{Arnab Dasgupta}
\email{arnabdasgupta@pitt.edu}
\altaffiliation{School of Liberal Arts, Seoul-Tech, Seoul 139-743, Korea}
\altaffiliation{PITT PACC, Department of Physics and Astronomy, University of Pittsburgh, Pittsburgh, PA 15260, USA}
\author{Takaaki Nomura}
\email{nomura@scu.edu.cn}
\altaffiliation{College of Physics, Sichuan University, Chengdu 610065, China}
\author{Hiroshi Okada}
\email{hiroshi.okada@apctp.org}
\altaffiliation{Asia Pacific Center for Theoretical Physics, Pohang 37673, Republic of Korea}
\altaffiliation{Department of Physics, Pohang University of Science and Technology, Pohang 37673, Republic of Korea}
\author{Oleg Popov}
\email{opopo001@ucr.edu}
\altaffiliation{Institute of Convergence Fundamental Studies, Seoul National University of Science and Technology, Seoul 139-743, Korea}
\altaffiliation{%
 Department of Physics, Korea Advanced Institute of Science and Technology, 291 Daehak-ro, Yuseong-gu, Daejeon 34141, Republic of Korea
}%
\altaffiliation{
 Scuola Normale Superiore, Piazza dei Cavalieri 7, 56126 Pisa, Italy
}%
\author{Morimitsu Tanimoto}
\email{tanimoto@muse.sc.niigata-u.ac.jp}
\altaffiliation{Department of Physics, Niigata University, Niigata 950-2181, Japan}
\date{\today}

\begin{abstract}
Minimalistic Dirac radiative neutrino mass model based on modular symmetry is proposed. We predict maximum number of observables possible including neutrino mass splittings, neutrino mass scale, lepton mixing angles, and Dirac phases in the leptonic sector with minimum number of input parameters possible. Model is capable of accommodating multicomponent dark matter, thanks to the $\emph{R}-$parity and accidental scotogenic $\mathbb{Z}_2$ discrete symmetry. Furthermore, even-though neutrinos are Dirac in our model, matter-antimatter asymmetry of the Universe is achieved via neutrinogenesis mechanism. Phenomenology of the dark sector including various dark matter candidates is briefly discussed.
\end{abstract}
\pacs{14.60.Pq, 95.35.+d, 12.60.-i, 14.60.St}
\keywords{neutrino mass, modular symmetry, flavor symmetry, dark matter}
\maketitle
%
\twocolumngrid
\begin{singlespace}
\tableofcontents
\end{singlespace}
\onecolumngrid
%
\section{Introduction}
\label{sec:intro}
Physics beyond the standard model (SM) is required in explaining some issues such as non-zero neutrino masses, existence of dark matter (DM) and matter-antimatter asymmetry of the universe.
In extending the SM, a new symmetry plays important roles to restrict structure of new physics which can realize, for example, 
stability of DM, neutrino mass generation at loop level forbidding tree level mass and origin of structure for neutrino mass matrix.
 Thus it is interesting to find a symmetry providing such properties with high predictability.
 
In controlling flavor structure, attractive framework of symmetries is proposed by papers~\cite{Feruglio:2017spp,
deAdelhartToorop:2011re}, in 2017, where they applied modular non-Abelian discrete flavor symmetries to quark and lepton sectors.
Remarkably this framework has advantage that any dimensionless couplings can also be transformed as non-trivial representations under those symmetries. 
As a result, we do not need copious scalars to find a predictive mass matrix. 
Furthermore we have a modular weight from the modular origin that can play a role in stabilizing DM when appropriate charge assignments are assigned to each of the fields in models.
Along the line of this idea, many approaches have appeared in the literature, {\it e.g.}, based on modular $A_4$~\cite{Feruglio:2017spp, Criado:2018thu, Kobayashi:2018scp, Okada:2018yrn, Nomura:2019jxj, Okada:2019uoy, deAnda:2018ecu, Novichkov:2018yse, Nomura:2019yft, Okada:2019mjf,Ding:2019zxk, Nomura:2019lnr,Kobayashi:2019xvz,Asaka:2019vev,Zhang:2019ngf, Ding:2019gof,Kobayashi:2019gtp,Nomura:2019xsb, Wang:2019xbo,Okada:2020dmb,Okada:2020rjb, Behera:2020lpd, Behera:2020sfe, Nomura:2020opk, Nomura:2020cog, Asaka:2020tmo, Okada:2020ukr, Nagao:2020snm, Okada:2020brs, Yao:2020qyy, Chen:2021zty, Kashav:2021zir, Okada:2021qdf, deMedeirosVarzielas:2021pug, Nomura:2021yjb, Hutauruk:2020xtk, Ding:2021eva, Nagao:2021rio,Okada:2021aoi},
$S_3$ \cite{Kobayashi:2018vbk, Kobayashi:2018wkl, Kobayashi:2019rzp, Okada:2019xqk, Mishra:2020gxg, Du:2020ylx},
$S_4$ \cite{Penedo:2018nmg, Novichkov:2018ovf, Kobayashi:2019mna, King:2019vhv, Okada:2019lzv, Criado:2019tzk,
Wang:2019ovr, Zhao:2021jxg, King:2021fhl, Ding:2021zbg, Zhang:2021olk, Qu:2021jdy,Nomura:2021ewm},
$A_5$~\cite{Novichkov:2018nkm, Ding:2019xna,Criado:2019tzk}, double covering of $A_5$~\cite{Wang:2020lxk, Yao:2020zml, Wang:2021mkw, Behera:2021eut}, larger groups~\cite{Baur:2019kwi}, multiple modular symmetries~\cite{deMedeirosVarzielas:2019cyj}, and double covering of $A_4$~\cite{Liu:2019khw, Chen:2020udk, Li:2021buv}, $S_4$~\cite{Novichkov:2020eep, Liu:2020akv}, and the other types of groups \cite{Kikuchi:2020nxn, Almumin:2021fbk, Ding:2021iqp, Feruglio:2021dte, Kikuchi:2021ogn, Novichkov:2021evw} in which masses, mixing, and CP phases for the quark and/or lepton have been predicted~\footnote{For interested readers, some literature reviews would be useful to understand the non-Abelian group and its applications to flavor structure~\cite{Altarelli:2010gt, Ishimori:2010au, Ishimori:2012zz, Hernandez:2012ra, King:2013eh, King:2014nza, King:2017guk, Petcov:2017ggy}.}. Majorana neutrino mass matrix with two texture zeros can be also realized applying  modular $\mathcal{A}_4$ symmetry~\cite{Zhang:2019ngf}. 
In addition to the lepton sector, stability of DM can be realized at fixed points under the modular $A_4$ symmetry~\cite{Kobayashi:2021ajl}.

Further researches are found; a systematic approach to understand the origin of CP transformations has been discussed in Ref.~\cite{Baur:2019iai}, CP/flavor violation in models with modular symmetry was discussed in Refs.~\cite{Kobayashi:2019uyt,Novichkov:2019sqv}, a possible correction from K\"ahler potential was discussed in Ref.~\cite{Chen:2019ewa}, and systematic analysis of the fixed points (stabilizers) has been discussed in Ref.~\cite{deMedeirosVarzielas:2020kji}.

In this paper, we construct a model realizing Dirac neutrino mass based on framework of modular $A_4$ symmetry with supersymmetry
in which we try to find minimal contents for new particles and modular forms. 
The manuscript is organized as follows: section~\ref{sec:model} describes the model at hand, generation of particle masses, including bosons and fermions, section~\ref{sec:lepto} describes the generation of the matter-anti-matter asymmetry of the universe with relevant constraints on the model, generation of the neutrino masses and other heavy neutral states is described in section~\ref{sec:mnu}, analysis and predictions for leptonic sector is executed in section~\ref{sec:neutrino_analysis}, section~\ref{sec:discussion} discusses possible dark matter candidates and various phenomenology of the model, section~\ref{sec:conclusion} concludes the work.

%
%

%
\section{Model}
\label{sec:model}
In this section we introduce our model that is SuperSymmetric (SUSY) and applies modular $A_4$ symmetry.
 Model contains no flavon fields and is build on Minimal SuperSymmetric Model (MSSM) by appending it with $\Bar{\nu}, \eta, \eta', \chi,$ and $N$ superfields. The particle content is given in Tab.~\ref{tab:particles} where we summarize representations under modular $A_4$, modular-weight $k$ and $3(B-L)$ value of each superfield. While $\bar{\nu}$ is added to make neutrinos Dirac, $\eta,\chi$, and $N$ superfields are needed for scotogenic (radiative) neutrino mass mechanism, where scotogenic symmetry is modular $\mathcal{A}_4$ symmetry together with $\mathcal{R}-$parity. Here $\eta'$ is added to cancel gauge anomaly since our model is SuperSymmetric. Model Lagrangian, \emph{aka} superpotential, is given as
\begin{align}
    \label{eq:lag}
    \mathcal{W} &= \bar{u}\pmb{y_u} Q H_u - \bar{d}\pmb{y_d} Q H_d + \mu H_u H_d \\
    &- \delta_1 L_1 (\pmb{y_e} \bar{e})_{1} H_d - \delta_2 L_{1^{\prime}} (\pmb{y_e} \bar{e})_{1^{\prime\prime}} H_d - \delta_3 L_{1^{\prime\prime}} (\pmb{y_e} \bar{e})_{1^{\prime}} H_d \nonumber \\
    &+ \alpha_1 L_1 (\pmb{y_l} N)_1 \eta + \alpha_2 L_{1^\prime} (\pmb{y_l} N)_{1^{\prime\prime}} \eta + \alpha_3 L_{1^{\prime\prime}} (\pmb{y_l} N)_{1^\prime} \eta \nonumber \\
    &+ \beta_1 (N \pmb{y_\nu})_{1} \Bar{\nu}_{1} \chi_1 + \beta_2 (N \pmb{y_\nu})_{1^{\prime}} \Bar{\nu}_{1^{\prime\prime}} \chi_1 + \beta_3 (N \pmb{y_\nu})_{1^{\prime\prime}} \Bar{\nu}_{1^{\prime}} \chi_1 \nonumber \\
    &+ \tilde \beta_1 (N \pmb{y_\nu})_{1} \Bar{\nu}_{1} \chi_2 + \tilde \beta_2 (N \pmb{y_\nu})_{1^{\prime}} \Bar{\nu}_{1^{\prime\prime}} \chi_2 + \tilde \beta_3 (N \pmb{y_\nu})_{1^{\prime\prime}} \Bar{\nu}_{1^{\prime}} \chi_2 \nonumber \\
    &+ \gamma_h \pmb{y_3} \eta \chi_1 H_d +\tilde \gamma_h \pmb{y_3} \eta \chi_2 H_d + \gamma_N \Lambda (N \pmb{y_n} N)_1 
    + \sum_{ij} \epsilon^{ij} \mu_\chi \pmb{y_\chi} \chi_i \chi_j + \text{H.c.} \, , \nonumber
\end{align}
where we are using a two component notation, following~\cite{Dreiner:2008tw}.
Here $\pmb{y_X} (\pmb{X}= e,l,\nu, 3, \chi, \chi^{\prime})$ denotes modular forms whose representations and corresponding modular weight are summarized in Table~\ref{tab:yukawas}.
We write modular forms $\pmb{y_3^{(2)}} = (y_{1},y_{2},y_{3})^T$, $\pmb{y_1^{(4)}} = y^2_1+2y_2y_3$ and $\pmb{y_3^{(4)}} = (y^2_1-y_2y_3, y^2_3-y_1y_2,y^2_2-y_1y_3)^T$, where $y_i$ is given by Dedekind eta$-$function $\eta(\tau)$ of modulus $\tau$ and its derivative $\eta'(\tau)$, as given in ref.~\cite{Feruglio:2017spp} ($y_i$ is written as $Y_i$ in the reference). On the other hand $\{\delta_a, \alpha_a, \beta_a, \tilde \beta_a,  \gamma_{h}, \tilde \gamma_h, \epsilon^{ij} \}$ are coupling constants.
Terms that are forbidden by various symmetries are
\begin{subequations}
\begin{align}
\label{eq:forb_pr}
    \mathcal{W}_{\not{P_R}} &= \mathcal{W}_{\not{P_R}}^{\text{MSSM}} + L H_u + \eta H_d + \bar{\nu}N + N\chi + \pmb{y_\chi^\prime} \chi \chi \chi, \\
\label{eq:forb_a4}
    \mathcal{W}_{\not{\mathcal{A}_4}} &= y_1^{(2)} L\bar{\nu} H_u + \Lambda y_1^{(2)} \bar{\nu} \bar{\nu} + y_{1,1',1''}^{(3)} \Lambda \eta L + y_1^{(3)} \Lambda \bar{\nu} \chi + y_3^{(1)} H_u H_d N, \\
\label{eq:forb_both}
    \mathcal{W}_{\not{P_R}\& \not{\mathcal{A}_4}} &= y_1^{(2)}\chi H_u H_d,
\end{align}
\end{subequations}
where matter parity and $\mathcal{R}-$parity are defined as $P_M = (-1)^{3(B-L)}$ and $P_{\mathcal{R}}=(-1)^{3(B-L)+2s}$ with $s$ being spin of the particle, respectively.
Parameters and observables of the leptonic sector are listed in Tab.~\ref{tab:parameters}.

If scalar $N$ somehow gets a non$-$zero VEV (it is $\mathcal{R}-$parity even), then neutrinos would get a tree$-$level (still Dirac) mass via Dirac seesaw (as explained in model$-$I tree level scenario of~\cite{Ma:2016mwh}). This does not happen in our case because $y_3^{(1)} H_u H_d N$(eq.~\ref{eq:forb_a4}) term is forbidden due to modular invariance of the super$-$potential. $y_3^{(1)} H_u H_d N$ is the only possible source term that can induce $\left\langle N\right\rangle\neq 0$. In other words, the VEV of $N$ is not induced by VEVs of $H_{u,d}$ and since $\mathcal{R}-$parity does not protect $\left\langle N\right\rangle=0$, it indicates that there must be another induced or accidental symmetry in the Lagrangian related to the fact that $\left\langle N \right\rangle=0$.
\\
The extra symmetry is $\mathbb{Z}_2$, which can be seen from the $\Lambda N \pmb{y_n} N$ term of eq.~\ref{eq:lag}. The $\mathbb{Z}_2-$odd particles under this accidental symmetry(this accidental symmetry is present because of modular $\mathcal{A}_4$ symmetry invariance of the superpotential) are $\hat{N},\hat{\chi},\hat{\eta}$. From this we can conclude that the lightest of these $P_\mathcal{R}-$even states is a dark matter (DM) candidate which makes our model a multi-particle dark matter model.

\begin{table}[H]
    \centering
    \begin{tabular}{ccccccc}
    \hline\hline
        S$-$Field & SU(3)$_c$ & SU(2)$_L$ & U(1)$_Y$ & $\mathcal{A}_4$ & $-k$ & $3(B-L)$ \\ \hline
        Q & {\bf 3} & {\bf 2} & $\frac{1}{6}$ & $1$ & $0$ & $1$ \\
        $\bar{u}$ & $\pmb{\bar{3}}$ & {\bf 1} & $-\frac{2}{3}$ & $1$ & $0$ & $-1$ \\
        $\bar{d}$ & $\pmb{\bar{3}}$ & {\bf 1} & $\frac{1}{3}$ & $1$ & $0$ & $-1$ \\
        L & {\bf 1} & {\bf 2} & $-\frac{1}{2}$ & $\pmb{1,1^{\prime},1^{\prime\prime}}$ & $-1$ & $-3$ \\
        $\bar{e}$ & {\bf 1} & {\bf 1} & $1$ & $\pmb{3}$
        & $-1$ & $3$ \\
        $\bar{\nu}$ & {\bf 1} & {\bf 1} & $0$ & $\pmb{1,1^{\prime},1^{\prime\prime}}$ & $-1$ & $3$ \\
        $N$ & {\bf 1} & {\bf 1} & $0$ & $\pmb{3}$ & $-1$ & $0$ \\
        $H_u$ & {\bf 1} & {\bf 2} & $\frac{1}{2}$ & $\pmb{1}$ & $0$ & $0$ \\
        $H_d$ & {\bf 1} & {\bf 2} & $-\frac{1}{2}$ & $\pmb{1}$ & $0$ & $0$ \\
        $\eta$ & {\bf 1} & {\bf 2} & $\frac{1}{2}$ & $\pmb{1}$ & $-2$ & $3$ \\
        $\eta'$ & {\bf 1} & {\bf 2} & $-\frac{1}{2}$ & $\pmb{1}$ & $-2$ & $3$ \\
        $\chi$ & {\bf 1} & {\bf 1} & $0$ & $\pmb{1}$ & $-2$ & $-3$ \\
    \hline\hline
    \end{tabular}
    \caption{Model particle content.\footnote{The matter$-$parity $(3(B-L))$ twin ($-$,$+$) particles are $(L,H_d),(\eta,H_u),(\Bar{\nu} \chi,N)$.} $k$ is the modular$-$weight.}
    \label{tab:particles}
\end{table}
%
\begin{table}[H]
    \centering
    \begin{tabular}{ccc}
    \hline\hline
        Field & $\mathcal{A}_4$ & $-k$ \\ \hline
        $\pmb{y_e}=\pmb{y_3^{(2)}}$ & $\pmb{3}$ & $2$ \\
        $\pmb{y_l}=\pmb{y_3^{(4)}}$ & $\pmb{3}$ & $4$ \\
        $\pmb{y_{\nu}}=\pmb{y_3^{(4)}}$ & $\pmb{3}$ & $4$ \\
        $\pmb{y_3}=\pmb{y_1^{(4)}}$ & $\pmb{1}$ & $4$ \\
        $\pmb{y_n}=\pmb{y_3^{(2)}}$ & $\pmb{3}$ & $2$ \\
        $\pmb{y_{\chi}}=\pmb{y_1^{(4)}}$ & $\pmb{1}$ & $4$ \\
        $\pmb{y_{\chi}^{\prime}}=\pmb{y_1^{(6)}}=6y_1y_2y_3$ & $\pmb{1}$ & $6$ \\
    \hline\hline
    \end{tabular}
    \caption{Modular transformations of Yukawas and dimensionfull parameters of the model.}
    \label{tab:yukawas}
\end{table}
\begin{table}[H]
    \centering
    \begin{tabular}{cc}
    \hline\hline
        Observable & Predicted/Input/Constrained \\ \hline
        $\Delta m^2_{\text{sol}}$ & P \\
        $\Delta m^2_{\text{atm}}$ & I \\
        $m_1$ & C \\
        $\sin^2\theta_{12}$ & P \\
        $\sin^2\theta_{23}$ & P \\
        $\sin^2\theta_{13}$ & P \\
        $\delta_{CP}$ & P \\
        $m_{ee}$ & P/C \\
        $\sum m_i$ & P \\
        $m_e,m_\mu,m_\tau$ & I \\
        \hline\hline
        Model parameter & Constrained by/Free \\ \hline
        $\pmb{y_e}$ & $\alpha_i,\tau$ \\
        $\pmb{y_{\nu}}$ & $\alpha_i,\tau$ \\
        $\tau$ & Scan \\
        $\mu_{H}$ & $m_{\nu}$ \\
        $\mu$ & $G_F,m_h, \frac{\partial V}{\partial H_u^0}=0, \frac{\partial V}{\partial H_d^0}=0$ \\
        $v_{u}, v_{d}\leftrightarrow v, \tan\beta$ & $m_e,m_\mu,m_\tau,G_F$ \\
        \hline\hline
    \end{tabular}
    \caption{Parameters and observables of the leptonic sector.}
    \label{tab:parameters}
\end{table}
%

\subsection{Boson sector}
\label{sec:boson}
Here we discuss mass eigenstates and mixings of neutral scalar bosons which are R-parity odd.
Because of the term $\gamma_h y_3\eta\chi_1 H_d +\tilde \gamma_h y_3\eta\chi_2 H_d$, the neutral components of inert bosons $\eta$ and $\chi_{1,2}$ mix with each other. We formulate their mixings as
\begin{align}
    \left(\begin{matrix}
    \chi_1^0   \\
    \chi_2^0 \\
     \eta_0    \\
      \end{matrix}\right) & =
  \begin{pmatrix}
1 &0&0 \\
0 &c_{H_{23}} &-s_{H_{23}} \\
0 &s_{H_{23}}  &c_{H_{23}} \\
\end{pmatrix}
\begin{pmatrix}
c_{H_{13}} & 0 &-s_{H_{13}} \\
0&1&0\\
s_{H_{13}} & 0& c_{H_{13}}\\
\end{pmatrix}
\begin{pmatrix}
c_{H_{12}} & -s_{H_{12}} &0\\
s_{H_{12}} & c_{H_{12}} &0\\
0&0&1 \\
\end{pmatrix}
       \left(\begin{matrix}
    H_1   \\
    H_2   \\
     H_3   \\
      \end{matrix}\right), \nonumber \\ 
      & \equiv U_H  \left(\begin{matrix}
    H_1   \\
    H_2   \\
     H_3   \\
      \end{matrix}\right),
\end{align}
where we consider mixing angles $s_{H_{ij}}\equiv \sin\theta_{H_{ij}}$,$c_{H_{ij}}\equiv \cos\theta_{H_{ij}}$, and mass eigenvalues $m_{H_{1,2,3}}$ as free parameters.
In this paper, we do not discuss charged scalar bosons since they are irrelevant for leptogenesis and neutrino mass generation.
Higgs sector in our model is the same as that of MSSM and we do not discuss here.
%
%
\subsection{Charged-lepton masses}
\label{sec:ell}
In this subsection we discuss charged lepton masses. 
Charged-lepton mass matrix is given through the following Lagrangian:
\begin{align}
  {\cal L}_\ell&=\delta_1 (\bar e_1y_1+\bar e_2 y_3+\bar e_3 y_2)H_d L_1
  +\delta_2(\bar e_2 y_2+\bar e_1 y_3+\bar e_3 y_1)H_d L_2 & \nonumber \\ 
  &+\delta_3 (\bar e_3 y_3+\bar e_1 y_2+\bar e_2 y_1)H_d L_3 + {\rm h.c.},
\end{align}
where $\pmb{y_3^{(2)}}\equiv [y_1,y_2,y_3]^T$ is applied for terms for second line in Eq.~\ref{eq:lag}.
After the EW spontaneously breaking, we find 
\begin{align}
m_\ell=\left[\frac{v_d}{\sqrt2}
\left(\begin{matrix}
    y_1 & y_3 & y_2  \\
     y_3 & y_2 & y_1   \\
      y_2 & y_1 & y_3   \\
      \end{matrix}\right)
      \left(\begin{matrix}
    \delta_1 & 0 & 0   \\
0 & \delta_2 & 0\\
0 & 0 & \delta_3   \\
      \end{matrix}\right)
      \right]_{\bar e L}.
    \end{align}
Then, the mass matrix is diagonalized by $D_\ell\equiv V_{e_R}^\dag m_\ell V_{e_L}$; $|D_\ell|^2=V_{e_L}^\dag m^\dag_\ell m_\ell V_{e_L}$.
%
\section{Leptogenesis}
\label{sec:lepto}
The resultant leptonic asymmetry is realised by satisfying the Sakharov's condition \cite{Sakharov:1967dj}. Now, in order to get a non-zero $CP$ violation one needs to satisfy the Nanopolous-Weinberg theorem \cite{Nanopoulos:1979gx} and along with that the condition pointed out by Adhikari-Rangarajan \cite{Adhikari:2001yr}. The first condition specifies atleast how many $B/L$ violating couplings on needs and the second condition tells us precisely where to keep place such couplings.

In our scenario the only particle which breakes the lepton-number is $\chi$ through it's mass term. The decay of $\chi$ will create an asymmetry in the right$-$handed sector. Then, eventually the asymmetry from the right$-$handed sector goes to the left$-$handed sector through the process $\bar{\nu} \widetilde{\chi} \rightarrow \nu^\dagger \widetilde{\eta}^\dagger$. The processes responsible for the generation of an asymmetry are given as follows:
\begin{figure}[H]
    \centering
    \begin{tabular}{lcr}
    \begin{tikzpicture}[/tikzfeynman/small]
        \begin{feynman}
        \vertex (i){$\chi$};
        \vertex [right = 2cm of i] (v1);
        \vertex [right = 1cm of v1] (v2);
        \vertex [above = 1cm of v2] (j){$N$};
        \vertex [below = 1cm of v2] (k){$\bar{\nu}$};
        \diagram*[small]{(i) -- [anti fermion](v1) -- [fermion](k),(v1) -- [charged scalar](j)};
        \end{feynman}
    \end{tikzpicture}&  
     \begin{tikzpicture}[/tikzfeynman/small]
        \begin{feynman}
        \vertex (i){$\chi$};
        \vertex [right = 2cm of i] (v1);
        \vertex [right = 1cm of v1] (v2);
        \vertex [above = 1cm of v2] (v3);
        \vertex [below = 1cm of v2] (v4);
        \vertex [right = 1cm of v3] (j){$N$};
        \vertex [right = 1cm of v4] (k){$\bar{\nu}$};
        \vertex[left = 0.5 of v3] (cv1);
        \vertex[below = 2 of cv1] (cv2);
        \diagram*[small]{(i) -- [fermion](v1) -- [anti fermion,edge label = $\bar{\nu}$](v3) -- [majorana, insertion=0.5, edge label = $\chi$](v4) -- [fermion](k),(v1) -- [anti charged scalar,edge label'=$N$](v4),(v3)--[charged scalar](j),(cv1) -- [scalar,style={thick,red}](cv2)};
        \end{feynman}
    \end{tikzpicture}&
    \begin{tikzpicture}[/tikzfeynman/small]
        \begin{feynman}
        \vertex (i){$\chi$};
        \vertex [right = 2cm of i] (v1);
        \vertex [right = 2cm of v1] (v3);
        \vertex [right = 2cm of v3] (v4);
        \vertex [right = 1cm of v4] (v5);
        \vertex [above = 1cm of v5] (j){$N$};
        \vertex [below = 1cm of v5] (k){$\bar{\nu}$};
        \vertex [above left = 1.8cm of v3] (cv1);
        \vertex [below = 2cm of cv1] (cv2);
        \diagram*[small]{(i) -- [fermion](v1) -- [anti fermion,half left,edge label = $\bar{\nu}$](v3)-- [majorana, insertion=0.5, edge label = $\chi$](v4)--[fermion](k),(v4) -- [charged scalar](j),(v1)--[anti charged scalar,edge label'= $N$](v3),(cv1)--[scalar,style={thick,red}](cv2)};
        \end{feynman}
    \end{tikzpicture}
    \end{tabular}
    \caption{The above figure shows the processes responsible for giving the $CP$ violation.}
    \label{fig:asym}
\end{figure}
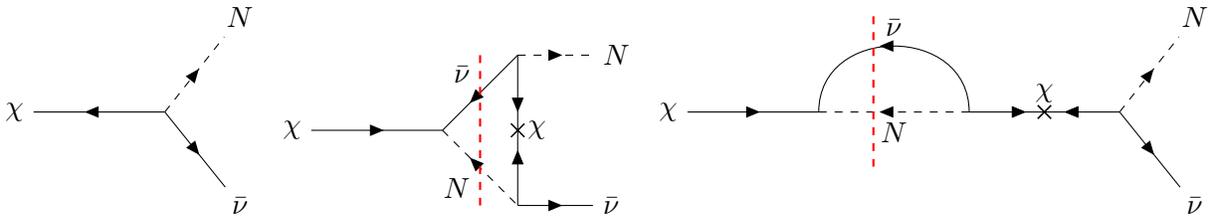
The diagram shown in fig.\ref{fig:asym} does gives a non-zero $CP$ which would eventually give an asymmetry in the right handed sector. The asymmetry then would be communicated to the left-handed sector through the following channel shown in fig \ref{fig:transfer}.
\begin{figure}[H]
    \centering
    \begin{tikzpicture}[/tikzfeynman/small]
        \begin{feynman}
        \vertex (i){$\tilde{\eta}^0$};
        \vertex [below = 2cm of i] (j){$\nu$};
        \vertex [below = 1cm of i] (v1);
        \vertex [right = 1cm of v1] (v2);
        \vertex [right = 2cm of v2] (v3);
        \vertex [right = 4cm of i] (k){$\widetilde{\chi}$};
        \vertex [right = 4cm of j] (l){$\bar{\nu}$};
        \diagram*[small]{(j) -- [anti fermion](v2) -- [majorana, insertion=0.5,edge label=$\widetilde{N} \quad \widetilde{N}$](v3) -- [fermion](l),(i)--[anti charged scalar](v2),(v3)--[charged scalar](k)};
        \end{feynman}
    \end{tikzpicture}
    \caption{The above process is responsible for transferring the asymmetry from the right handed sector to the left handed sector. }
    \label{fig:transfer}
\end{figure}
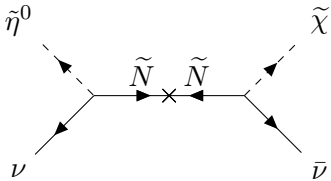
Although the above process would act as a source term and as well as the major wash-out channel. So, in order to get the most asymmetry one would require this wash-out factor to go out-of-equilibrium at the earlier time. Which boils down in satisfying the following relation
\begin{align}
    \frac{\gamma^{eq}_{Scatt}(\widetilde{\chi}\bar{\nu} \rightarrow \tilde{\eta}^0\nu )}{Hs} < 1.
\end{align}
In order to get the resultant asymmetry we need to solve the coupled boltzmann equations.
\begin{subequations}
\begin{align}
    \frac{dY_\chi}{dz} &= \frac{-1}{zsH}\left[\left(\frac{Y_\chi}{Y^{eq}_\chi} - 1\right)\gamma_D(\chi \rightarrow \nu_R \widetilde{N}) + \left(\frac{Y^2_\chi}{(Y^{eq}_\chi)^2}-1\right)\gamma^{eq}_{Scatt}(\chi \chi \rightarrow all)\right], \\
    \frac{dY_{\Delta R}}{dz} &= \frac{1}{zsH}\left[\left(\frac{Y_\chi}{Y^{eq}_\chi} - 1\right)\epsilon \gamma_D(\chi \rightarrow \nu_R \widetilde{N}) - \frac{Y_{\Delta R}}{Y^{eq}_l}\gamma_D(\chi \rightarrow \nu_R \widetilde{N}) \right. \\
    &- \left.2 \frac{Y_{\Delta R}}{Y^{eq}_l}\left[\gamma^{eq}_{scatt}(\nu_R \widetilde{N}\rightarrow \bar{\nu}_R \widetilde{N}) 
    + \gamma^{eq}_{scatt}(\nu_R \nu_R\rightarrow \widetilde{N} \widetilde{N})\right] \right. \nonumber \\
    &+ \left. \left(\frac{Y_{\Delta L} - Y_{\Delta R}}{Y^{eq}_l}\right)\gamma^{eq}_{Scatt}(\widetilde{\chi}\nu_R \rightarrow H\nu_L )\right], \nonumber \\
    \frac{dY_{\Delta L}}{dz} &= \frac{1}{zsH}\left[\left(\frac{Y_{\Delta R} - Y_{\Delta L}}{Y^{eq}_l}\right)\gamma^{eq}_{Scatt}(\widetilde{\chi}\nu_R \rightarrow H\nu_L )\right],
\end{align}
\end{subequations}
where $z=M_\chi/T$, $\gamma_D(i_1 \rightarrow f_1 + f_2 + \cdots)$ and $\gamma^{eq}_{scatt}(i_1 i_2\rightarrow f_1 + f_2 + \cdots)$ is given as,
\begin{subequations}
\begin{align}
    \gamma_D(i_1 \rightarrow f_1 + f_2 + \cdots) &= \frac{g_i}{2\pi^2}m^2_{i_1}TK_1(m_{i_1}/T)\Gamma(i_1 \rightarrow f_1 + f_2 + \cdots), \\
    \gamma^{eq}_{Scatt}(i_1 i_2\rightarrow f_1 + f_2 + \cdots) &= \frac{g_{i_1}g_{i_2}T}{8\pi^4}\int^\infty_{s_{in}}ds \frac{p_{in}p_{out}}{\sqrt{s}}|\mathcal{M}(i_1 i_2 \rightarrow f_1 + f_2 + \cdots)|^2K_1(\sqrt{s}/T),
\end{align}
\end{subequations}
in which $K_{1}$ is the modified Bessel function. As to get the estimate of the asymmetry we start by calculating the CP asymmetry parameter $\varepsilon$ for $\chi \rightarrow N\nu_R$ decays which is given as 
\begin{subequations}
\begin{align}
    \varepsilon_i &= \frac{1}{8\pi (Y_\nu^\dagger Y_\nu)_{ii}}\Im[(Y^\dagger_\nu Y_\nu)^2_{ij}]\frac{1}{\sqrt{x_{ji}}}\mathcal{F}(x_{ji}),  \\
    \mathcal{F}(x_{ji}) &= \sqrt{x_{ji}}\left[f(x_{ji}) - \frac{\sqrt{x_{ji}}}{x_{ji}-1} \right], \\
    f(x_{ji}) &= \sqrt{x_{ji}}\left[1 + (1+x_{ji})\ln \left(\frac{x_{ji}}{x_{ji}+1}\right)\right],
\end{align}
\end{subequations}
with $x_{ji} = M^2_j/M^2_i$ and $(Y_\nu)_{1(2)} \equiv \beta_a \pmb{y_\nu} (\tilde \beta_a \pmb{y_\nu})$ omitting flavor indices. Furthermore, the decay $\Gamma_i$ is given as 
\begin{align}
    \Gamma_i &= \frac{M_{\chi_i}}{8\pi}(Y^\dagger_\nu Y_\nu)_{ii}. 
\end{align}
Now assuming only the resonant case and taking the mass differences of the decaying $\chi$ masses to be $M_{\chi_j} - M_{\chi_i} = \Gamma_i/2$. Which simplifies the total asymmetry as follows 
\begin{align}
    \varepsilon &= \sum_i \varepsilon_i = \sin(2\phi).
\end{align}
If we demand the out-of-equilibrium to occur around $T\sim M_\chi$ we can safely put a constraint on $y_\chi$'s by the following relation
\begin{subequations}
\begin{align}
    \frac{\Gamma_{\chi}}{H} &= \left(\frac{8\pi^3g_*}{90}\right)^{-1/2}\frac{M_{pl}}{M_\chi}\frac{(Y^\dagger_\nu Y_\nu)_{11}}{8\pi} = 1 \quad \textrm{assuming} \quad M_{\chi_1}>M_{\chi_2}, \\
    (Y_{\nu})^2_1 &= \frac{M_{\chi_1}}{M_{pl}}\sqrt{\frac{8\pi^3g_*}{90}} \label{eq:ynu}, \\
    (Y_{\nu})^2_1 &= 1.43\times 10^{-15}\left(\frac{M_{\chi_1}}{1 {\rm TeV}}\right) = (Y_{\nu})^2_2 \label{eq:condition},
\end{align}
\end{subequations}
Now, in order to get the estimate we assume the processes leading to left-right equilibration include $\widetilde{\eta}\nu_L \rightarrow \widetilde{\chi}\nu_R$ mediated with s-channel exchange of an $\widetilde{N}$. Approximately, at high temperatures these processes have a rate
\begin{align}
    \Gamma_{L-R} &\sim 2\frac{|(Y_{\nu})_1|^2|Y_l|^2}{M^4_{\widetilde{N}}}T^5,
\end{align}
where $Y_l \equiv \alpha_a \pmb{y_l}$.
This should be compared with the Hubble rate
\begin{align}
    H&= \sqrt{\frac{8\pi^3g_*}{90}}\frac{T^2}{M_{pl}}.
\end{align}
The considerable constraint will be coming from the highest temperatures when $T \simeq M_{\chi}$, i.e. those at which the asymmetry is generated
\begin{align}
    2\frac{|(Y_{\nu})_1|^2|Y_l|^2}{M_{\widetilde{N}}}M^3_{\chi_1} &\lesssim \frac{1}{M_{pl}}\sqrt{\frac{8\pi^3g_*}{90}}.
\end{align}
The ratio boils down to
\begin{subequations}
\begin{align}
    |Y_l|^2 &\lesssim \frac{M^4_{\widetilde{N}}}{2 M^4_{\chi_1}} |Y_l|^2 \leq 4\pi, \label{eq:yl} \\
    M_{\widetilde{N}} &= (8\pi)^{1/4} M_{\chi_1},
\end{align}
\end{subequations}
which basically tells us that the ratio $M_{\widetilde{N}}\sim M_{\chi_1}$. Hence, the final asymmetry can be given as 
\begin{subequations}
\begin{align}
    \eta_B &= a_{sph}\frac{86}{2387}\varepsilon Y^{eq}_{\chi_1}(z=1),  \\
    \eta_B &= 4.479\times 10^{-5} \sin (2\phi),
\end{align}
\end{subequations}
where $a_{sph} = 28/79$ and the observed baryonic asymmetry $\eta^{obs}_B = 6\times 10^{-10}$ which can be transformed to 
\begin{align}
    \sin (2\phi) &= 1.34\times 10^{-5} \quad \sim \phi = 6.7\times 10^{-6}.
\end{align}
%

%
%
\section{Neutrino masses}
\label{sec:mnu}
Here generation of the Dirac neutrino masses is discussed, but we start with the masses of the heavy neutral fermions.

\subsection{Heavy neutral masses}
Before formulating the active neutrino mass matrix, 
let us formulate the heavier Majorana neutral fermion $N$.
The explicit form of Lagrangian is found as
\begin{align}
    {\cal L}_N&=M_0\left[ y_1(2N_1N_1-N_2N_3-N_3N_2)
    + y_2(2N_2N_2-N_1N_3-N_3N_1)\right. \nonumber\\
&\left. + y_3(2N_3N_3-N_1N_2-N_2N_1)\right] +{\rm h.c.}
\end{align}{}
Thus, the Majorana mass matrix is give by
\begin{align}
M_N =M_0
\left(\begin{matrix}
    2 y_1 & - y_3 & - y_2   \\
     - y_3& 2 y_2 & - y_1   \\
      - y_2  & - y_1  & 2 y_3   \\
      \end{matrix}\right).
      \end{align}
Then, this is diagonalized by $D_N\equiv U^T M_N U$; $|D_N|^2\equiv U^\dag M^\dag_N M_N U$, furthermore $N\equiv U\psi$, where $\psi$ is mass eigenstate of $N$.
\subsection{Neutrino mass generation}
\begin{figure}[H]
    \centering
    \includegraphics[width=0.7\textwidth]{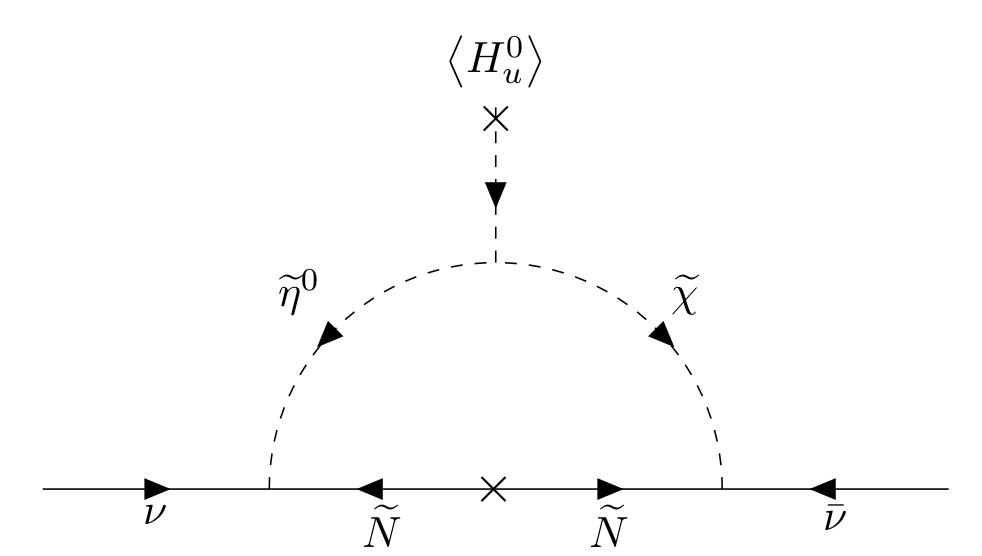}
    \caption{Dirac scotogenic neutrino mass diagram}
    \label{fig:mnua4dirac2}
\end{figure}

In our model tree level Dirac neutrino mass term is fobidden by modular invariance of the superpotential, while scotogenic Dirac neutrino mass term is allowed due to presence of Majorana fermion $N$.
The relevant interactions to generate Dirac neutrino mass matrix is given by
\small{
\begin{align}
  {\cal L}_\nu&=\alpha_1 (N_1^T y'_1+ N_2^T y'_3+ N_3^T y'_2)\eta L_1
  +\alpha_2( N_2^T y'_2+ N_1^T y'_3+ N_3^T y'_1)\eta L_2
    +\alpha_3( N_3^T y'_3+ N_1^T y'_2+ N_2^T y'_1)\eta L_3\nonumber\\
    &
    +\beta_1 \bar\nu_1( N_1 y'_1 + N_2 y'_3+ N_3 y'_2)\chi_1
  +\beta_2 \bar\nu_2( N_2 y'_2+ N_1 y'_3+ N_3 y'_1)\chi_1
    +\beta_3 \bar\nu_3( N_3 y'_3+ N_1 y'_2+ N_2 y'_1)\chi_1
   + {\rm h.c.}\nonumber\\
       &
    +\tilde \beta_1 \bar\nu_1( N_1 y'_1 + N_2 y'_3+ N_3 y'_2)\chi_2
  +\tilde \beta_2 \bar\nu_2( N_2 y'_2+ N_1 y'_3+ N_3 y'_1)\chi_2
    +\tilde \beta_3 \bar\nu_3( N_3 y'_3+ N_1 y'_2+ N_2 y'_1)\chi_2
   + {\rm h.c.}, \nonumber\\
&   \supset N^T y_\eta \nu\eta^0 + \bar\nu y_\chi N \chi^0_1 + \bar\nu \tilde y_\chi N \chi^0_2 + {\rm h.c.}, \nonumber\\
& =\psi^T U^T y_\eta \nu ((U_{H})_{31} H_1+(U_{H})_{32} H_2 + (U_{H})_{33} H_3) + \bar\nu y_\chi U \psi ((U_{H})_{11} H_1+(U_{H})_{12} H_2 + (U_{H})_{13} H_3) \nonumber \\
& + \bar\nu \tilde y_\chi U \psi ((U_{H})_{21} H_1+(U_{H})_{22} H_2 + (U_{H})_{23} H_3) + {\rm h.c.},
   \end{align}
}   
where ${y_3^{(4)}}\equiv [y_1',y_2',y_3']^T$, and we rewrite the interaction with mass eigenvector in the last line.
Then, we find each of Yukawa matrix to be
\begin{align}
y_\eta = \alpha_1 \tilde y_\eta & = \alpha_1 \left[
\left(\begin{matrix}
    y_1' & y_3' & y_2'   \\
     y_3' & y_2' & y_1'   \\
      y_2' & y_1' & y_3'   \\
      \end{matrix}\right)
      \left(\begin{matrix}
    1 & 0 & 0  \\
0 & \tilde \alpha_2 & 0\\
0 & 0 & \tilde \alpha_3   \\
      \end{matrix}\right)
      \right]_{N^T L}, \\
     y_\chi (\tilde y_\chi) & = \left[
\left(\begin{matrix}
    y_1' & y_3' & y_2'  \\
     y_3' & y_2' & y_1'  \\
      y_2' & y_1' & y_3'  \\
      \end{matrix}\right)
      \left(\begin{matrix}
    \beta_1 (\tilde \beta_1) & 0 & 0   \\
0 & \beta_2 (\tilde \beta_2) & 0\\
0 & 0 & \beta_3 (\tilde \beta_3)   \\
      \end{matrix}\right)
      \right]_{\bar \nu N},
    \end{align}
where $\tilde{\alpha}_{2,3} \equiv \alpha_{2,3}/\alpha_1$ and $\alpha_1$ is factored out for convinience in numerical analysis.
In terms of these interactions Dirac scotogenic neutrino mass diagram is given in Fig.~\ref{fig:mnua4dirac2}.
Analytic form of the mass matrix is estimated to be
\begin{subequations}
\begin{align}
&  m_{\nu_{ij}} = - \frac{\alpha_1}{(4 \pi)^2} \sum_{a=1}^3 \sum_{A=1}^3 (U^T \tilde y_\eta)_{ia} D_{N_a} \left[ (y_\chi U)_{aj} (U_H)_{3 A} (U_H)_{1 A}  + (\tilde y_\chi U)_{aj} (U_H)_{3 A} (U_H)_{2 A} \right]f (r^a_A), \\
& f(r^a_A) = \frac{r^a_A\ln r^a_A}{1-r_A^a},
\end{align}
\end{subequations}
where $r^a_{A}\equiv \frac{m_{H_{A}}^2}{D_a^2}$.
Here, we redefine $\tilde m_\nu\equiv \frac{m_\nu}{\alpha_1}$.
Then the neutrino mass matrix is diagonalized by $U_\nu^T \tilde m_\nu U_\nu\equiv$diag.($\tilde m_1,\tilde m_2,\tilde m_3$).
Finally, we find
\begin{subequations}
\begin{align}
 &   \alpha_1^2 =\frac{\Delta m^2_{\rm atm}}{\tilde m^2_3-\tilde m^2_1},\quad \Delta m^2_{\rm sol}=\frac{\tilde m^2_2 -\tilde m^2_1}{\tilde m^2_3-\tilde m^2_1}\Delta m^2_{\rm atm},\quad ({\rm NH}),\\
&    \alpha_1^2 =\frac{\Delta m^2_{\rm atm}}{\tilde m^2_2 -\tilde m^2_3},\quad \Delta m^2_{\rm sol}=\frac{\tilde m^2_2 -\tilde m^2_1}{\tilde m^2_2-\tilde m^2_3}\Delta m^2_{\rm atm},\quad ({\rm IH}),
\end{align}
\end{subequations}
where we require $ \alpha_1^2\le 4 \pi$ to guarantee perturbativity of the Yukawa coupling.
Then, one finds $U_{PMNS}=V^\dag_{eL} U_\nu$.
Each of mixing is given in terms of the component of $U_{MNS}$ as follows:
\begin{align}
\sin^2\theta_{13}=|(U_{PMNS})_{13}|^2,\quad 
\sin^2\theta_{23}=\frac{|(U_{PMNS})_{23}|^2}{1-|(U_{PMNS})_{13}|^2},\quad 
\sin^2\theta_{12}=\frac{|(U_{PMNS})_{12}|^2}{1-|(U_{PMNS})_{13}|^2}.
\end{align} 
%
%
Neutral fermion mass matrices are given by
\begin{align}
\centering
\label{eq:nu_even_odd}
    &\left(\begin{matrix}\begin{matrix}
    0 & m_d \\ m_d & {\color{blue}0} 
    \end{matrix} &
    \Large{{\color{blue}\pmb{0}}_{\mathcal{\not{A}}_4}} \\
    \Large{{\color{blue}\pmb{0}}_{\mathcal{\not{A}}_4}} & \begin{matrix} 0 & \pmb{y_3} v_d \\ \pmb{y_3} v_d & \varepsilon \mu_\chi \pmb{y_\chi} \end{matrix}
    \end{matrix}\right)
    &\left(\begin{matrix}\begin{matrix}
    0 & \mu \\ \mu & 0
    \end{matrix} &
    \Large{{\color{blue}\pmb{0}}_{\mathcal{\not{A}}_4}} \\
    \Large{{\color{blue}\pmb{0}}_{\mathcal{\not{A}}_4}} & \gamma_N \Lambda \pmb{y_n}
    \end{matrix}\right)
\end{align}
in the $(\nu_L, \bar{\nu}_L, \eta_L, \chi_L)(P_{\mathcal{R}}=+)$ and $(\Tilde{h}_{uL}^0,\Tilde{h}_{dL}^0,\Tilde{N}_L)(P_{\mathcal{R}}=-)$ basis, respectively, where bar over $\nu$ is notational like in~\cite{Martin:1997ns}. Entries indicated in dark blue are forbidden by $\mathcal{A}_4$ modular invariance. Therefore, we conclude that neutrinos are of Dirac type and do not mix with other neutral fermions due to modular $\mathcal{A}_4$ symmetry and $P_\mathcal{R}$ parity conservation.

Considering the constraints from eq.\eqref{eq:ynu} for $(Y_{\nu})_{1,2}$ and also assuming $M_{\widetilde{N}}= M_{\chi_1}$ we have $\widetilde{y}_l \lesssim 1/\sqrt{2}$ from eq. \eqref{eq:yl}. Now, assuming the mass of $\chi$'s to be $\mathcal{O}(1)$ TeV, $Y_{\nu} \simeq 3.78\times 10^{-8}$ to obtain observed baryon asymmetry. 
We thus chose small $\beta_a$ and $\tilde \beta_a$ values in our numerical calculation below to achieve condition for $(Y_\nu)_{1(2)} \equiv \beta_a \pmb{y_\nu} (\tilde \beta_a \pmb{y_\nu})$.




%
\section{Neutrino analysis and discussion}
\label{sec:neutrino_analysis}
In this section, we perform numerical $\Delta \chi^2$ analysis searching for allowed region, satisfying neutrino oscillation data and LFVs. Also we show our predictions, where we apply the best fit values for charged-lepton masses.
Here, we concentrate on the NH case, since IH is disfavored which would be clarified by analytical estimation as can be seen in the previous section(if possible).

\begin{figure}[h!]
    \centering
    \includegraphics[width=0.7\textwidth]{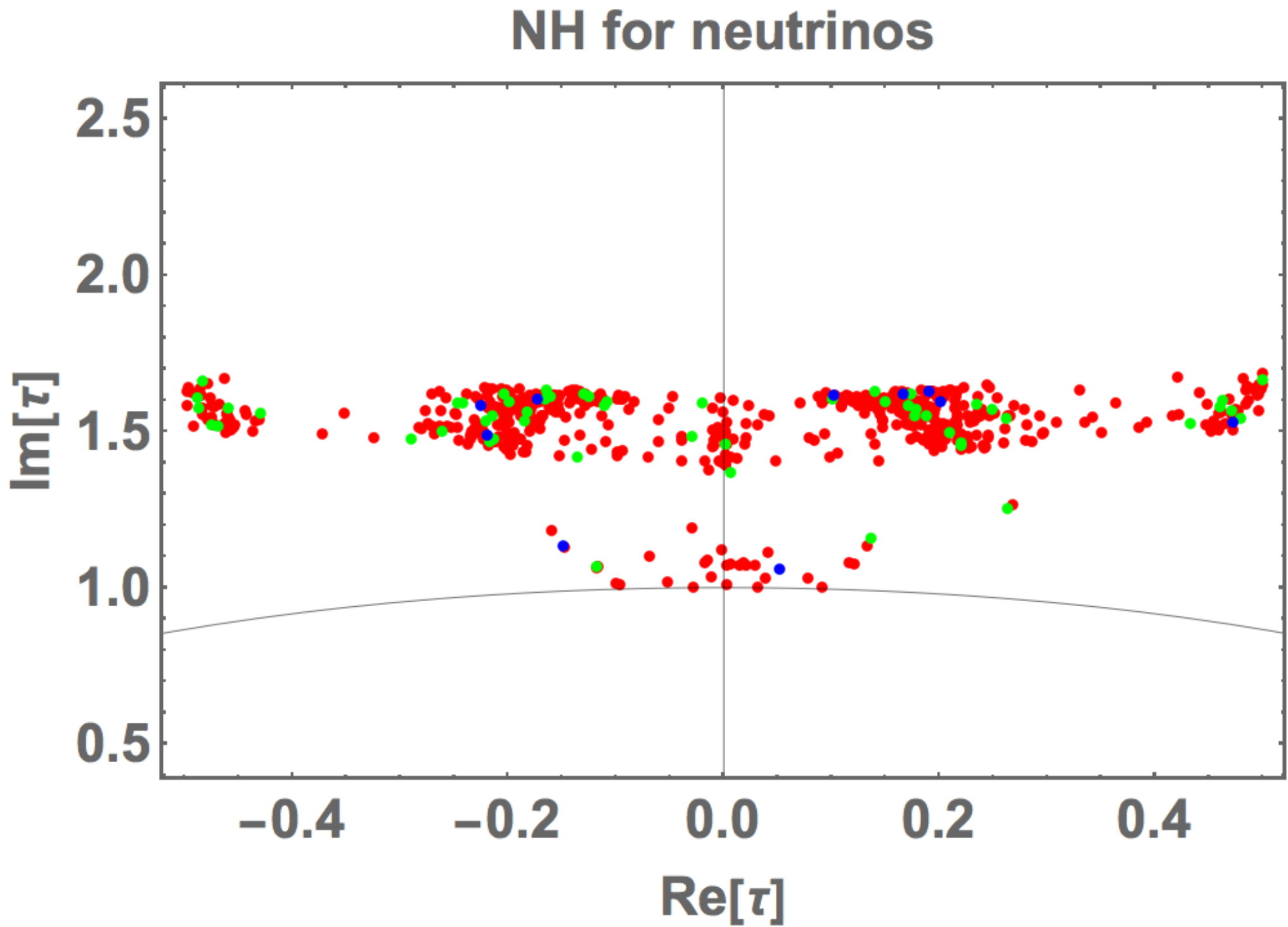}
    \caption{Allowed region of modulus $\tau$, where the blue points are allowed within 2, green ones within 3, and red one within 5 of $\Delta \chi^2$ analysis.}
    \label{fig:tau_nh}
\end{figure}

In our numerical analysis, we randomly scan free parameters in the  following ranges 
\begin{align}
& \{ \tilde\alpha_2,  \tilde\alpha_3\} \in [10^{-4},1.0],\quad \{ \beta_{1,2,3},  \tilde\beta_{1,2,3}\} \in [10^{-10}, 10^{-6}], \quad \sin \theta_{H_{12,13,23}} \in [-0.5, 0.5], \nonumber \\
& \{ m_{H_{1}},  M_0\} \in [10^3,10^4]\ {\rm GeV}, \quad m_{H_2} \in [1.0, 1.1] M_0, \quad m_{H_3} \in [1.1, 1.2] M_0,
\end{align}
where $\tau$ runs over the fundamental region.
Here we choose small $\beta_a(\tilde \beta_a)$ values that are required to realize Baryon assymmetry indicated by Eq.~\ref{eq:condition}. 
Then, we perform numerical analysis and discuss below.
In Fig.~\ref{fig:tau_nh}, we shows the allowed region between real part of $\tau$ and imaginary part of $\tau$, where the blue points are allowed within 2, green ones within 3, and red one within 5 of $\sqrt{\Delta \chi^2}$
for five accurately known observables $\Delta m^2_{\rm atm},\Delta m^2_{\rm sol},s_{12}^2,s_{23}^2,s_{13}^2$ in Nufit 5.0~\cite{Esteban:2018azc, Esteban:2020cvm, Nufit:2029abc}.
The real part of $\tau$ runs whole the range in the fundamental region, while the imaginary one runs over the region of $[1.0-1.7]$.

\begin{figure}[tb!]\begin{center}
\includegraphics[width=0.7\textwidth]{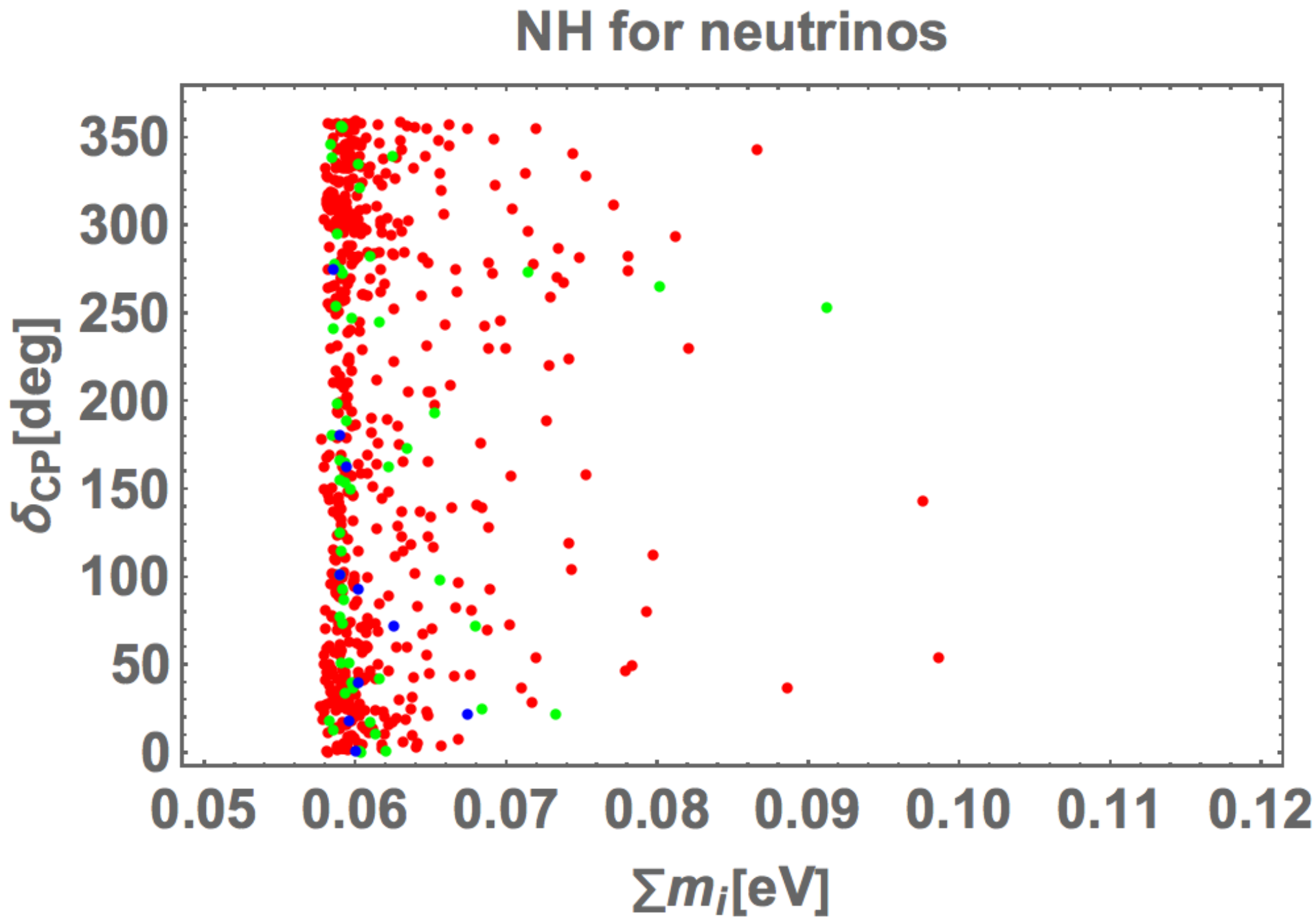}
\caption{Correlation between the sum of neutrino mass eigenvalues ($\sum m_i$ eV) and Dirac CP phase $\delta_{CP}$, where the legend is the same as the case of Fig.1.}   
\label{fig:sum-dcp_nh}\end{center}\end{figure}

Fig.~\ref{fig:sum-dcp_nh} demonstrates the correlation between the sum of neutrino mass eigenvalues ($\sum m_i$ eV) and Dirac CP phase $\delta_{CP}$. The legend is the same as Fig.1.
Dirac CP runs whole the ranges, while $\sum m_i$ tends to be localized at around $0.06$ eV. It implies that the lightest neutrino mass is very small compared to the other two masses.

\begin{figure}[tb!]\begin{center}
\includegraphics[width=0.45\textwidth]{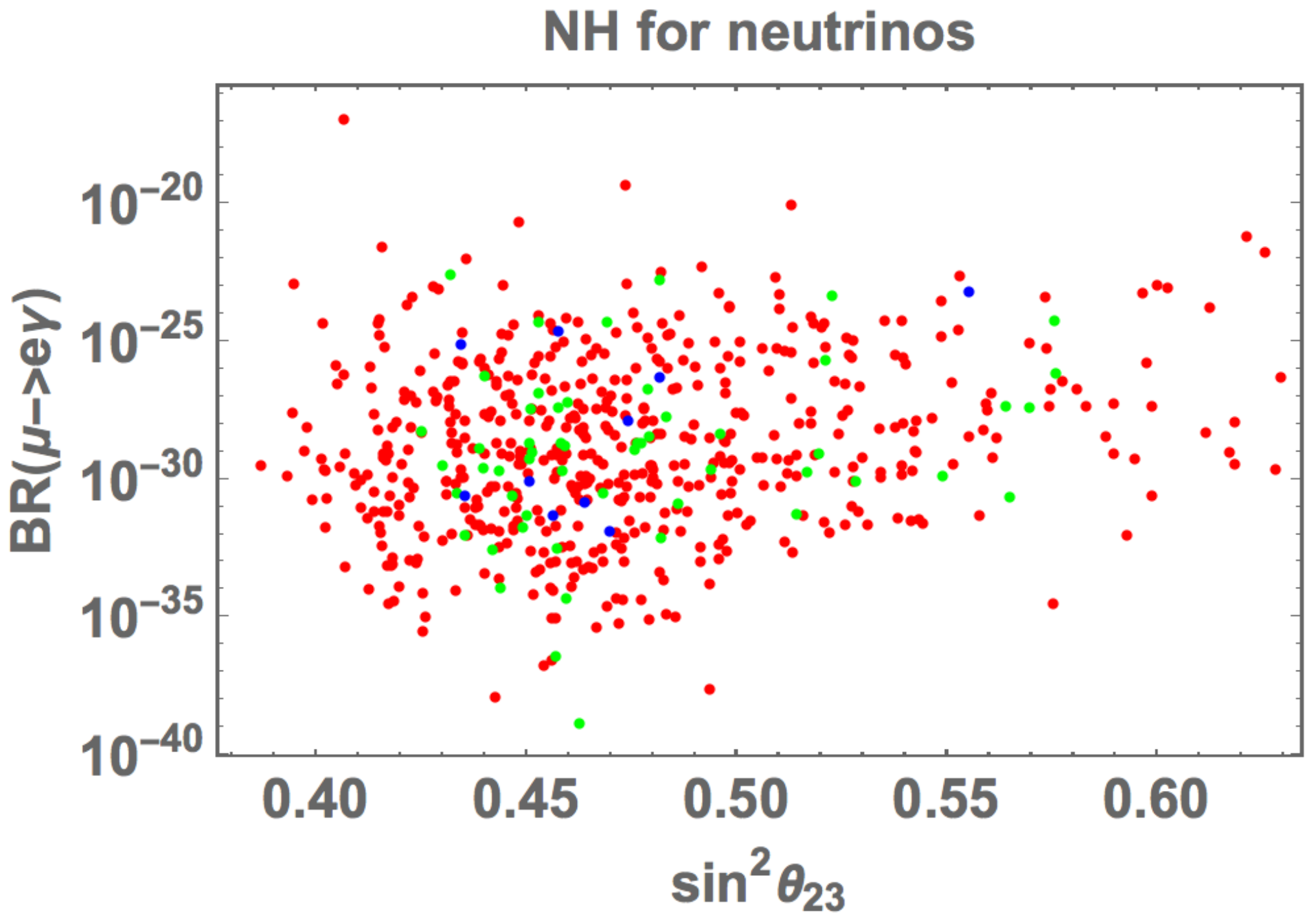}\
\includegraphics[width=0.45\textwidth]{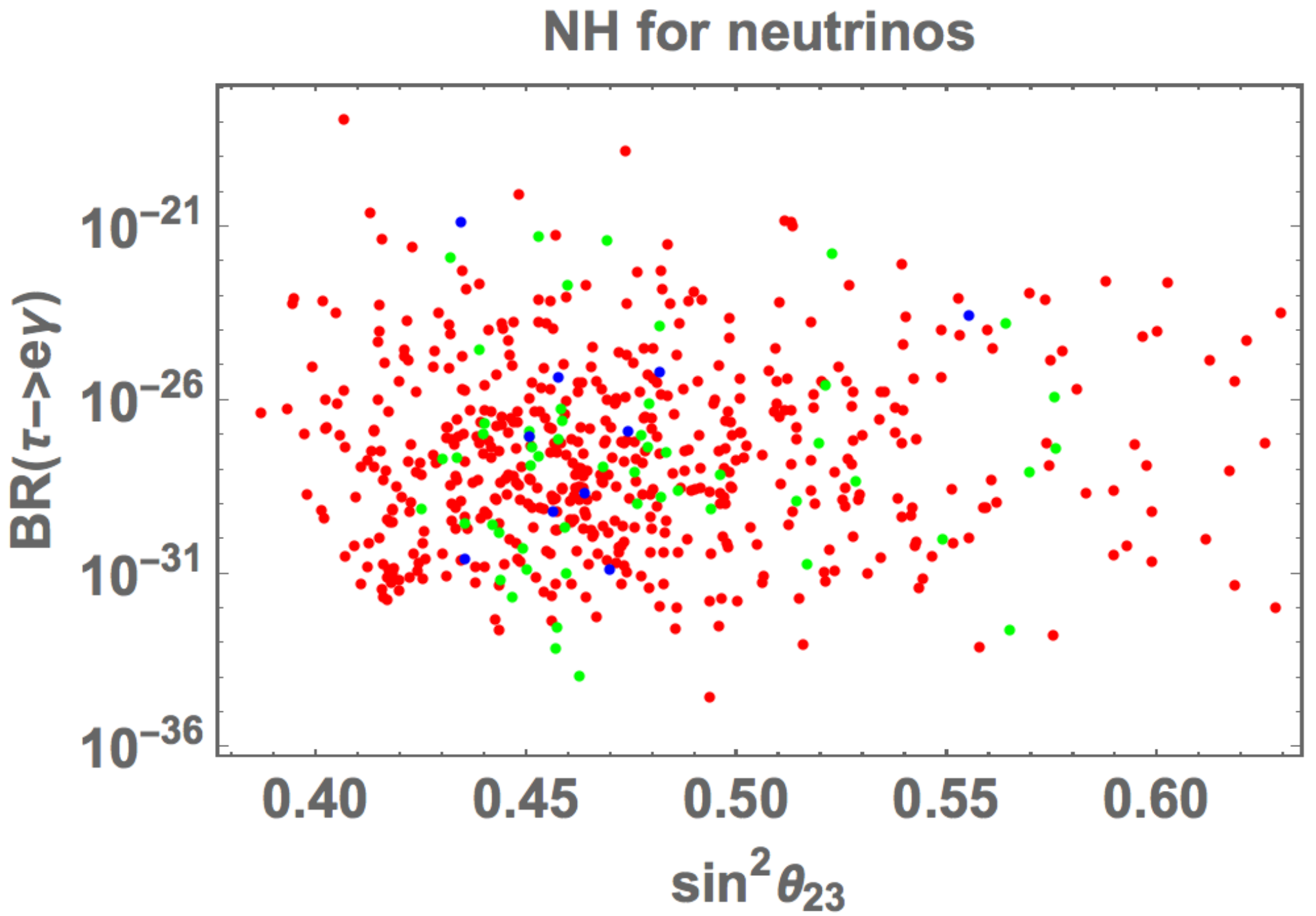}\
\includegraphics[width=0.45\textwidth]{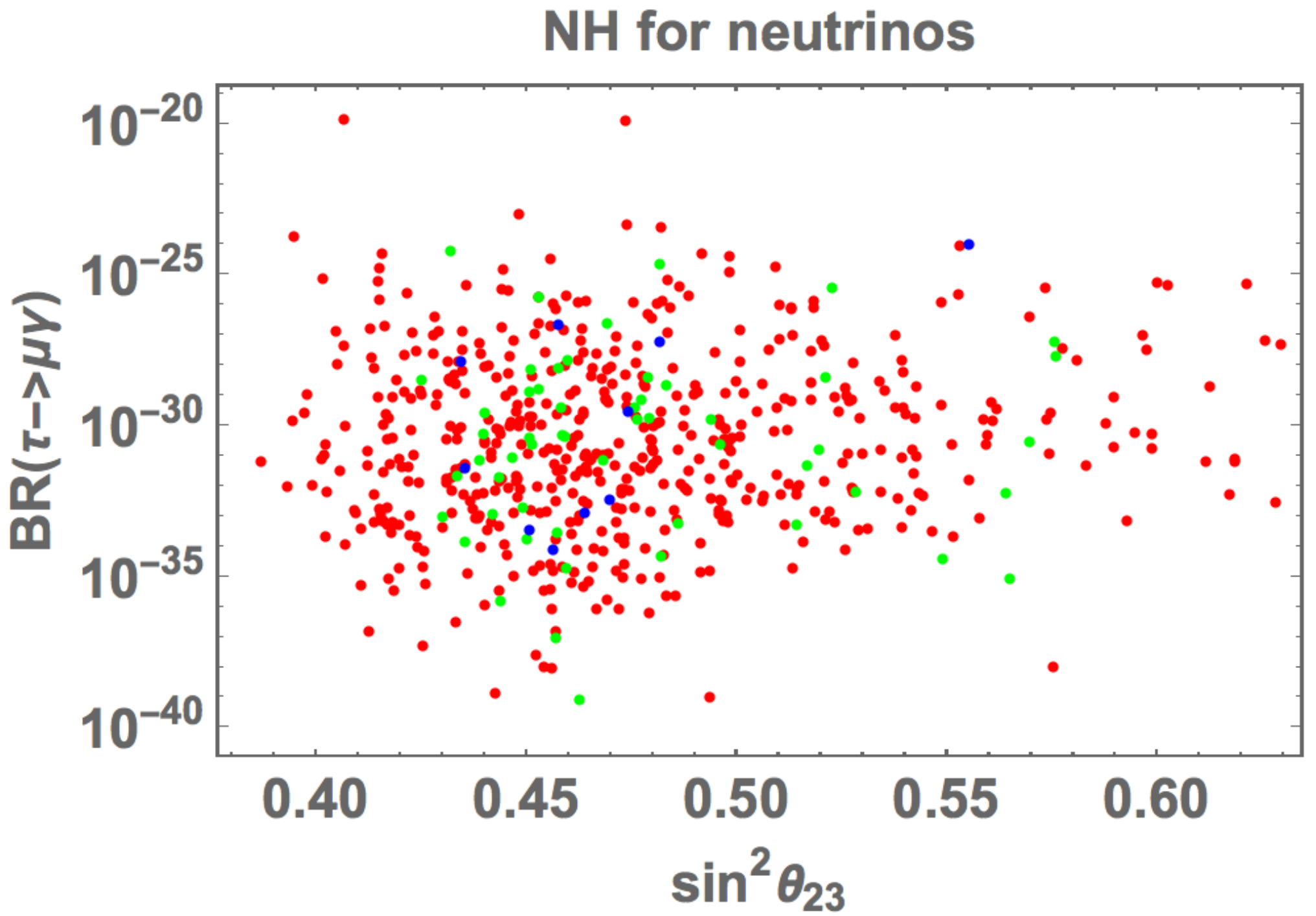}
\caption{Relations between LFVs and $s^2_{23}$, where the legend is the same as the case of Fig.1. }   
\label{fig:lfvs_nh}\end{center}\end{figure}

Figs.~\ref{fig:lfvs_nh} show relations between LFVs and $s^2_{23}$. All the three BR's of LFVs are less than $10^{-19}$,
which are much smaller than the current upper experimental bounds.

%
Finally, we show a benchmark in table~\ref{bp-tab_nh}, where we select it so that $\sqrt{\Delta \chi^2}$ is minimum.
The mass matrices for dimensionless neutrino and charged-lepton are found as
\begin{align}
V_{e_L}&=
\left[\begin{array}{ccc} 
-0.200 & 0.959 & 0.200  \\
-0.0596 + 0.0129 i& 0.187 - 0.0407 i & -0.957 +  0.208 i\\
-0.955 - 0.211 i& -0.203 - 0.0450 i & 0.0197 + 0.00429 i \\
 \end{array}\right],\\
 U_\nu&=
\left[\begin{array}{ccc} 
0.338 & 0.730 & -0.594  \\
-0.206 - 0.132 i& -0.355 - 0.412 i& -0.554 - 0.582 i \\
-0.906 - 0.0648 i& 0.408 + 0.0710 i& -0.0147 + 0.0505 i \\
 \end{array}\right].
 \end{align}
%
\begin{table}[h!]
	\centering
	\begin{tabular}{|c|c|} \hline 
			\rule[14pt]{0pt}{0pt}
		$\tau$ & $0.102244 + 1.61566 i$     \\ \hline
		\rule[14pt]{0pt}{0pt}
		$[s_{H_{12}},s_{H_{23}},s_{{31}}]$ & $[0.393, 0.0975, 0.00692]$     \\ \hline
		\rule[14pt]{0pt}{0pt}
		$[\alpha_\ell,\beta_\ell,\gamma_\ell]$ & $[0.0596, 0.000298, 0.979]$     \\ \hline
		\rule[14pt]{0pt}{0pt}
				$[\alpha_1,\tilde\alpha_2,\tilde\alpha_3]$ & $[6.64\times10^{-7}, 0.00966 + 0.000711 i, 0.00272 + 0.758 i]$     \\ \hline
		\rule[14pt]{0pt}{0pt}
%
		$[\beta_1,\beta_2,\beta_3]$ & $[(17.9 - 3.90 i)\times10^{-10} , (16.1 - 2.82 i)\times10^{-9} , (208 + 2.83 i )\times10^{-9} ]$ \\ \hline
		\rule[14pt]{0pt}{0pt}
		$[\tilde\beta_1,\tilde\beta_2,\tilde\beta_3]$ & $[ (7.42 + 5.70 i ) \times10^{-10} , (175 - 2.74 i)\times10^{-9}, (769 + 9.91 i ) \times10^{-10} ]$ \\ \hline
		\rule[14pt]{0pt}{0pt}
		$[m_{H_1},m_{H_2},m_{H_3}]$ & $[4661, 4871, 5493]\ {\rm GeV}$  \\ \hline
		\rule[14pt]{0pt}{0pt}
		%
		$[M_{0},D_{N_1},D_{N_2},D_{N_3}]$ & $[8100, 6340, 10062, 16402]\ {\rm GeV}$  \\ \hline
		\rule[14pt]{0pt}{0pt}
		$\Delta m^2_{\rm atm}$  &  $2.52\times10^{-3} {\rm eV}^2$    \\ \hline
		\rule[14pt]{0pt}{0pt}
		$\Delta m^2_{\rm sol}$ &  $7.53\times10^{-5} {\rm eV}^2$       \\ \hline
		\rule[14pt]{0pt}{0pt}
		$\sin^2\theta_{12}$  & $ 0.295$ \\ \hline
		\rule[14pt]{0pt}{0pt}
		$\sin^2\theta_{23}$  &  $ 0.451$    \\ \hline
		\rule[14pt]{0pt}{0pt}
		$\sin^2\theta_{13}$  &  $ 0.0219$   \\ \hline
		\rule[14pt]{0pt}{0pt}
		$\delta_{CP} $ &  $18.0^\circ$    \\ \hline
		\rule[14pt]{0pt}{0pt}
		$\sum m_i$ &  $59.0$\,meV    \\ \hline
		\rule[14pt]{0pt}{0pt}
		$\sqrt{\Delta\chi^2}$ &  $1.32 $   \\ \hline
		\hline
	\end{tabular}
	\caption{A benchmark point of our input parameters and observables, where we select it so that $\sqrt{\Delta \chi^2}$ is minimum.}
	\label{bp-tab_nh}
\end{table}

%
%
%
\section{Discussion}
\label{sec:discussion}
Our model is capable of accommodating multiparticle dark matter (DM) due to presence of $P_\mathcal{R}$ and modular $\mathcal{A}_4$ symmetry invariances. A $P_\mathcal{R}$ \emph{odd} candidate ($\widetilde{N}_L,\widetilde{\eta}_s,\widetilde{\chi}_s$) is the canonical SUSY WIMP, which in our model is connected to the radiative neutrino mass generation via scotogenic mechanism. Second component of the multiparticle DM is guaranteed by the accidental $\mathbb{Z}_2$ discrete symmetry which is induced by the modular $\mathcal{A}_4$ invariance of the model. Here possible DM candidates are odd under this $\mathbb{Z}_2$ symmetry and even under $P_\mathcal{R}$, which are $N_s,\eta^0_L,\chi_L$ superfields.
Furthermore, in our scenario we are assuming $\chi_L$ to be heavier in order to generate asymmetry in the right-handed sector. The mixture of $\eta^0_L,\chi_L$ (as shown in eq.~\ref{eq:nu_even_odd}) is not singlet under SM gauge symmetry therefore $N_s$, which is a SM singlet, is the best DM candidate odd under accidental $\mathbb{Z}_2$ symmetry. Further details on dark matter abundance and direct detection are left for elsewhere.
%
\section{Conclusion}
\label{sec:conclusion}
In this work a Dirac radiative neutrino mass model based on modular $\mathcal{A}_4$ symmetry was presented. Being a scotogenic neutrino mass model, it demonstrates a natural connection between naturally small Dirac neutrino mass origin and existence of dark matter. Modular $\mathcal{A}_4$ symmetry in this work achieves simultaneously three main goals: Firstly, the Diracness of the neutrinos, \emph{aka} forbids all majorana neutrino mass terms, secondly it builds a connection of neutrino with dark matter through the well know scotogenic mechanism, and finally, it predicts and reproduces Dirac phase as well as neutrino mass splittings in the leptonic sector. This model favors normal neutrino mass hierarchy, while disfavors the inverted one. Heavy neutral, dark, fermions are of the order of $\mathcal{O}(1-10)$ TeV, dark neutral scalars are of the order $\mathcal{O}(1-5)$~TeV, whereas light neutrino masses are of the order of $0.1$ eV, and the sum of neutrino masses is around $60$~meV. Even though, neutrinos are Dirac and the lepton number is conserved in our model, we achieve the matter asymmetry of the universe by means of the lepton number violation in the right$-$handed neutrino sector via a mechanism known as \emph{neutrinogenesis}. The required phase in the combination of the lepton sector yukawa couplings is of the order $\mathcal{O}(10^{-6})$. Model is able to predict preferred Dirac phase in the leptonic sector, as well as neutrino mass splittings and mass order, accommodate multicomponent dark matter, thanks to the \emph{R}$-$parity and accidental scotogenic $\mathbb{Z}_2$ symmetry with minimum set of input parameters. This is the first model of Dirac scotogenic neutrino mass based on modular symmetries.
%
\acknowledgments
This research was supported by an appointment to the JRG Program at the APCTP through the Science and Technology Promotion Fund and Lottery Fund of the Korean Government. This was also supported by the Korean Local Governments - Gyeongsangbuk-do Province and Pohang City (H.O.). A.D. and O.P. are supported by the National Research Foundation of Korea Grants No. 2017K1A3A7A09016430 and No. 2017R1A2B4006338. OP is also supported by the Samsung Science and Technology Foundation under Grant No. SSTF-BA1602-04 and National Research Foundation of Korea under Grant Number 2018R1A2B6007000. Feynman diagrams were created using Ti\emph{k}Z-Feynman package~\cite{Ellis:2016jkw}.
\appendix
% ...
%
\bibliography{references}
\end{document}